\documentclass[12pt]{JHEP3}
\pdfoutput=1
\usepackage{amsmath,amssymb}
\usepackage{yfonts}

\usepackage[pdftex]{graphicx}

\newcommand{\be}{\begin{equation}}
\newcommand{\ee}{\end{equation}}
\newcommand{\beqa}{\begin{eqnarray}}
\newcommand{\eeqa}{\end{eqnarray}}

\def\med{\frac{1}{2}}

\def\im{\mathop{\rm Im}}
\def\d{\partial}
\def\dA{{\cal A}}

\newcommand{\qn}{\textswab{q}}
\newcommand{\wn}{\textswab{w}}

\newcommand{\Dis}{{\tilde{d}}}
\newcommand{\Diff}{D}

\def\med{\frac{1}{2}}
\def\d{\partial}

\relax

\title{A note on conductivity and charge diffusion in holographic flavor systems}

\author{ Javier Mas\footnote{javier.mas@usc.es}, Jonathan P. Shock\footnote{shock@fpaxp1.usc.es}
and Javier Tarr\'\i o\footnote{tarrio@fpaxp1.usc.es}
\\
 Departamento de F\'\i sica de Part\'\i culas,
Universidade de Santiago de Compostela \\ 
and\\
Instituto Galego de F\'\i sica de Altas Enerx\'\i as (IGFAE)\\

E-15782 Santiago
de Compostela, Spain\\
}

\abstract{ We analyze the  charge diffusion and conductivity in a 
$Dp/Dq$ holographic setup that is dual to a supersymmetric Yang-Mills theory  in $p+1$ dimensions with  $N_f\ll N_c$ flavor  degrees of freedom at finite temperature and nonvanishing $U(1)$ baryon number chemical potential. We provide a new derivation of the results that generalize the membrane paradigm to the present context. We perform a  numerical analysis in the particular case of the $D3/D7$ flavor system. The results obtained  support the validity of the Einstein relation  at finite chemical potential. 
}
 
\begin{document}

\section{Introduction}

The robustness of the renowned result $\eta/s = 1/4\pi$  is  widely thought to be related to universal properties of black hole horizons.  For $p+1$-dimensional conformal field theories  with a conserved charge that admit a dual holographic description a similar relation has been proposed  for the  diffusion constant, $\Diff$, \cite{Kovtun:2008kx}. Recently, some
understanding of how both results stated above could be derived from gravitational data that live at the black hole horizon has  been pursued. In \cite{Kovtun:2003wp}, by relating the charge diffusion in the boundary field theory  to the analog process that occurs on the ``stretched horizon"  in the so called membrane paradigm,  a closed analytic expression for the diffusion constant, $\Diff$, was obtained as a a product of two factors. A local one, evaluated at the horizon, and a non-local one, involving an integral along the full range of the holographic radial direction. Certainly this smelled of holography, a point that  was clarified later  in  \cite{Starinets:2008fb},  where the same formula was derived from a purely AdS/CFT construction.
More recently  \cite{Myers:2007we} this factorized structure for the diffusion constant, was seen to match with the Einstein relation $\Diff = \sigma \chi^{-1}$. In it, the conductivity $\sigma$ was given by the local factor. The other one, namely the  integral along $r$, corresponded
exactly to the inverse of the charge susceptibility  $\chi^{-1}$, defined in equilibrium thermodynamics as the response of the charge density $n_q$ to a change in chemical potential $\mu$
\be
\chi =\left. \frac{\d n_q}{\d \mu} \right\vert_{T} \, .\label{defsuscep}
\ee
A holographic model where we can compute the above quantities in a controlled manner is naturally given by a set of $N_f$ ``flavor" probe Dq branes, placed in the gravitational black hole background created by $N_c \gg N_f$ ``color"  Dp branes. In order to model a  chemical potential of the boundary field theory
the world-volume $U(1)$ gauge field $A_t$ on the probe branes has to be switched on \cite{Nakamura:2006xk,Kobayashi:2006sb}.
These background ingredients are enough to compute the susceptibility   (\ref{defsuscep}).
On the other hand, both  the conductivity and the diffusion constant are transport coefficients, and their computation typically involves fluctuations.
From the point of view of the underlying QFT, $\sigma$ and $\Diff$ are calculated from different channels of the retarded 2-point function of a conserved current, the transverse and longitudinal channels respectively. Up to Lorentz structures, the retarded current-current 2-point function is expressed in terms of  two scalar functions $\Pi_\perp(k)$ and $\Pi_{||}(k)$. 
Fick's law for the conserved charge density $j^0$ implies the existence of a universal  pole in the hydrodynamic limit $k\to 0$  in $\Pi_{||}(k)$ from where the diffusive dispersion relation can be read off
\be
w =  - i \Diff q^2 + {\cal O}(q^3)\, . 
\ee
On the other hand, from linear response theory we can calculate the conductivity from the
following static limit \footnote{For an electromagnetic conductivity, we should supplement
this expression with a factor $e^2$ of the electromagnetic coupling. The same factor would show up in 
the susceptibility (see discussion in \cite{Kovtun:2008kx}) and they would cancel out in the diffusion constant. We will simply set $e=1$.)} (see \cite{Kovtun:2008kx,CaronHuot:2006te}).
\be
\sigma = -  \lim_{w\to 0} \frac{{\rm Im}\Pi_{\perp}(\omega)}{\omega}\, ,
\ee
The fact that these three quantities obey the Einstein relation $\Diff\chi = \sigma$ provides a non-trivial consistency check of the hydrodynamic picture of holography. Also the probe approximation is consistent with it. In fact,  the parametric scaling
$N_f \ll N_c$  implicit in the probe approximation is correctly accounted for. Both
$\chi$ and $\sigma$ arise from normalized quantities that ultimately derive from the action, and therefore show the correct scaling $N_f N_c\ll N_c^2$ with respect to bulk quantities. On the contrary $\Diff$ stems from a pole condition and, at probe-level,  does not scale with $N_f$. Backreaction should account for corrections of order $N_f/N_c$. This similar to the  situation considered in  \cite{Mateos:2006yd} for the quotient $\eta/s$

The present paper is organized as follows. In section \ref{holoset} we  establish the class of models we shall be working with. In  section \ref{seccon} we will obtain an expression for the
conductivity which includes and generalizes the one recently found in \cite{Iqbal:2008by}. 
Next, in section \ref{secsusc}, we study the charge susceptibility $\chi$, for which we also find an integral expression involving background fields that naturally extend the $n_q= 0$ case to non-zero baryon number. 
Finally, in \ref{secdifu}, we turn to the diffusion constant, $\Diff$. First we reanalyze the
case of zero baryon number, following a simplified version of the argument in \cite{Starinets:2008fb}. After that, we turn to the full system with chemical potential. The main difficulty stems from the fact that, in computing $\Pi_{||}$, longitudinal fluctuations mix up with scalar perturbations of the probe brane profile. We prove that, unlike the case $n_q=0$, the diffusion constant in general is not given by the one found in \cite{Kovtun:2003wp} using the membrane paradigm. So far we are able to provide a closed expression for $\Diff$ in terms of fluctuations, though not purely  in terms of background quantities. Nevertheless, in section \ref{secd3d7} we present numerical evidence that
the Einstein relation holds also for $n_q \neq 0$. Assuming this is also true in general, the upshot
is that the diffusion constant $\Diff$ can be computed from background fields as $\sigma\chi^{-1}$.

\section{The holographic setup \label{holoset}}
In this section we shall define the relevant notation. We will have in mind a generic holographic construction, where $N_f$ Dq branes are embedded in the ambient metric of $N_c\gg N_f$ Dp branes in the quenched approximation, where no backreaction of the flavor branes  is taken into account. Therefore we shall parameterize a 10-dimensional metric in the form
\be
ds^2=g_{00} (r) {\rm d}x_0^2 + g_{ii} (r) {\rm d} \vec{x}_p^2 + g_{rr} (r) {\rm d}r^2 + g_{\theta\theta} (r) {\rm d}\Omega_n^2 + g_{\psi\psi} (r) {\rm d}\psi^2 + g_{\varphi\varphi} (r) {\rm d}\Omega_{7-p-n}^2 \, ,
\ee
written to simplify the notation of a Dp/Dq intersection where the  probe Dq branes wrap an $n$-sphere and the transverse space is spanned by  coordinates $\psi$, $\varphi^i$.  The presence of a horizon  in the background metric at $r=r_H$ will be encoded in  the following relations
\beqa
g_{00}&=&-g_{ii}(r) f(r)\, , \nonumber\\
g_{rr}&=&G(r) f(r)^{-1}\, ,\nonumber
\eeqa
and $f(r) = (r-r_H) F(r)$ with $F(r)$ analytic at $r=r_H$.
The  embedding of the probe brane and the $U(1)$ gauge field strength are parametrized by functions $\psi(r)$ and $A_\mu(r)$ respectively, whose equations of motion are derived  from the Born-Infeld action 
\be
{\cal S}_{DBI} = - N_f T_{Dq}  \int d^{q+1}\xi e^{-\phi}\sqrt{ - \det(g_{ab} + 2 \pi \alpha' F_{ab}) }\, .
\ee
The Wess-Zumino term plays no role in the following. Here we have introduced the tensor
\be
\gamma_{ab}=g_{ab}+ 2\pi\alpha' F_{ab}\, ,
\ee
and we shall be interested in solutions that involve a non-zero value for the temporal component
 of the gauge field
\be
\gamma_{0r}=  2\pi\alpha' A_t'(r) =- \gamma_{r0}\, ,
\label{gaugepot}
\ee 
corresponding to a non-zero chemical potential. 
For a nonvanishing value of the gauge potential as in (\ref{gaugepot}) the matrix $\gamma_{ab}$ is
non-diagonal, so we list here the inverse components $\left(\gamma^{ab}\gamma_{ca}=\delta^b_c\right)$
\beqa
&&\gamma_{00}=-\gamma_{ii} f(r) ~~;~~ \gamma_{rr}=G(r) f(r)^{-1} ~~;~~ f(r)=(r-r_H) F(r)\, , \nonumber\\
&&\gamma^{00}=\frac{\gamma_{rr}}{\gamma_{00}\gamma_{rr}+(\gamma_{0r})^2} ~~;~~\gamma^{ii}=\frac{1}{\gamma_{ii}}\, ~~;~~\gamma^{\theta\theta}=\frac{1}{\gamma_{\theta\theta}}, \nonumber \\
&&\gamma^{rr}= \frac{\gamma_{00}}{\gamma_{00}\gamma_{rr}+(\gamma_{0r})^2}~~;~~\gamma^{0r}= -\gamma^{r0}=\frac{-\gamma_{0r}}{\gamma_{00}\gamma_{rr}+(\gamma_{0r})^2} \nonumber \, ,
\eeqa
where $\gamma_{ii}$, $\gamma_{\theta\theta}$, $\gamma_{0r}$ and $G(r)$ are regular at the horizon.
The temperature can be expressed in terms of these coefficients in the near horizon limit as
\be
T = \left. \frac{1}{4\pi} \frac{ \gamma_{ii} f'}{\sqrt{-\gamma_{00} \gamma_{rr}}} \right\vert_{r\to r_H}\, .
\label{temphawk}
\ee
In general, these coefficients depend on the radial coordinate through the  background profile $\psi(r)$ and the world-volume gauge field $A_\mu(r)$.
Reduced to such degrees of freedom, and integrated along the internal unit $n$-sphere of volume $\Omega_n$, the
$DBI$ action acquires the form
\be
{\cal S}_{DBI} = 
  - N_f T_{Dq}  \Omega_n \int d^{q+1-n} \xi\,   e^{-\phi}
\sqrt{ -\left( \gamma_{00} \gamma_{rr} + \gamma_{0r}^2 \right)  \gamma_{ii}^p \gamma_{\theta\theta}^n } 
\, ,
\ee
  The charge density $n_q$ can be obtained
from the electric displacement  \cite{Kobayashi:2006sb}
\be
n_q = \frac{\delta {\cal S}_{DBI}}{\delta A_t'(r)}  =  \frac{\cal N}{2\pi\alpha'} \, e^{-\phi}\sqrt{-\gamma}  \gamma^{0r}\, ,\label{betadis}
\ee
with $ \gamma = \det \gamma_{ab}$ without the angular components of the unit $n$-sphere and  ${\cal N} =N_f T_{D_q}(2\pi\alpha')^2 \Omega_n$.
From the equation of motion for $A_t$, it follows that the charge density is independent of the radial direction, $\d_r n_q = 0$.

 In the following sections we shall deal with the computation of retarded correlators. For this we consider fluctuations in the worldvolume fields of the following form
\beqa
\psi(r,x) &\to &~ \psi(r) + \epsilon e^{-i(\omega x^0 - q x^1)} \Psi(r)\, , \nonumber\\
A_\mu(r,x) &\to & A_\mu(r) + \epsilon e^{-i(\omega x^0 - q x^1)}\dA_\mu (r)\, .
\eeqa
If we expand the Dirac-Born-Infeld lagrangian to second order  in powers of $\epsilon$
\be
{\cal L} = {\cal L}_0 + \epsilon {\cal L}_1+ \epsilon^2 {\cal L}_2 + ...~,
\ee
upon imposing the equations of motion for the background fields,  the linear term, ${\cal L}_1$, vanishes identically. The linearized equations for the perturbations  $\Psi$ and $\dA$ are derived from the quadratic piece.
In ref. \cite{Mas:2008jz} a detailed analysis of the retarded correlators was performed. We refer the reader to that reference for details on the equations of motion and the spectral functions. 

\section{The conductivity \label{seccon}}

The electrical DC conductivity may be obtained from the zero-frequency slope of the trace of the spectral function \cite{CaronHuot:2006te,Mateos:2007yp}
\be
\sigma = \left. \frac{1}{2(p-1)} \lim_{\omega\to 0}\frac{\chi^\mu{_\mu}(k)}{\omega} 
\right\vert_{\omega = |{\bf q}|} = 
\frac{1}{2p} \lim_{\omega\to 0}\frac{\chi^\mu{_\mu}(\omega, {\bf q} = 0)}{\omega} \, .
\label{condchi}
\ee
The spectral function is split into two orthogonal components 
(see \cite{Myers:2007we,Mas:2008jz} for notation and conventions)
\be
\chi^\mu{_\mu}(k) = -2(p-1) {\rm Im} \Pi^\perp(k) - 2 {\rm Im} \Pi^{||}(k)  \, .\label{trspden} 
\ee
Taking into account the boundary conditions for the scalar function $ \Pi^\perp(k)$ and $ \Pi^{||}(k)$ 
\be
\Pi_\perp(\omega, {\bf q}=0 ) = \Pi_{||}(\omega, {\bf q}=0 )  ~~~;~~~\Pi_{||}(\omega = \pm |{\bf q}|) = 0\, .
\ee
From  (\ref{condchi}), the conductivity is uniquely obtained from the following expression
\be
\sigma= -    \lim_{\omega\to 0}\frac{\im\Pi_\perp(\omega)}{\omega}\, ,
\ee
independent of ${\bf q}$ (we shall confirm this independence shortly).
In order to compute $\Pi_\perp$ we  make use of the Minkowskian prescription given in
\cite{Son:2002sd} (see also \cite{Kovtun:2005ev}).  To start with, we must solve the equations of motion for the transverse electric field fluctuations  $E_\perp = \omega \dA_\perp$ with $\dA_\perp = \dA_{2,3}$.
\be
E_\perp'' + \partial_r \log\left[ e^{-\phi}\sqrt{-\gamma} \gamma ^{ii} \gamma ^{rr} \right] E_\perp' - \frac{\omega^2 \gamma ^{00} + q^2 \gamma ^{ii}}{\gamma ^{rr}}E_\perp = 0\, .
  \label{transEOM}
\ee
The horizon $r=r_H$ is a regular singular point, and so we may perform a Frobenius analysis. Introducing as usual the dimensionless quantities $\wn = \omega/(2\pi T)$ and $\qn=q/(2\pi T)$, the indicial exponents are found to be $\zeta_\pm = \pm i\wn/2 $.
We select  $\zeta_- $ corresponding to incoming wave boundary conditions and set
\be
E_\perp = f^{-i \frac{\wn}{2} } E_{\perp,reg}\, .
\ee
In order to analyze the solution to (\ref{transEOM}) in the hydrodynamic limit, a power counting parameter $\lambda$  is introduced. Setting $\wn\rightarrow \lambda \wn\,  , ~ \qn \rightarrow \lambda \qn$
as well as
\be
E_\perp \rightarrow f^{-i\lambda\frac{\wn}{2}}\left(E_{\perp,reg}^{(0)} + \lambda E_{\perp,reg}^{(1)} + ...\right)= E_{\perp,reg}^{(0)} + \lambda 
\left( E_{\perp,reg}^{(1)} -i\frac{\wn}{2}E_{\perp,reg}^{(0)} \log f  \right)+ ... 
\ee
 Expanding   (\ref{transEOM}) in powers of $\lambda$ one easily solves  for 
$ E_{\perp,reg}^{(0)}(r) $ in closed form
\be 
E_{\perp,reg}^{(0)}(r) = C_1+C_2\int_{r_H}^r\frac{dr }{e^{-\phi}\sqrt{-\gamma}\gamma^{ii}\gamma^{rr}}\, .
\label{ordzero}
\ee
The integrand diverges as a pole near the horizon. This enforces $C_2 = 0$ and we set $C_1=1$ which is simply an overall normalisation. This can then be used to solve to next order  $\lambda$ in closed form 
\be
E_{\perp,reg}^{(1)}(r) = C_4 \int_{C_3}^r\frac{ {\rm d}r }{e^{-\phi}\sqrt{-\gamma}\gamma^{ii}\gamma^{rr}} +i \frac{\wn}{2}\log f(r )\, .
\ee
 The logarithmic divergence at the horizon must be cancelled among the two contributions to $E_\perp^{(1)}(r) $ by suitably tuning $C_4 \equiv i\wn C$, where
\be
C =\left.  -\frac{1}{2} e^{-\phi} \sqrt{-\gamma} \gamma^{rr} \gamma^{00} f' \right\vert_{r\to r_H}\, .
 \label{constantC4}
\ee
The finite part is fixed by adjusting $C_3$ so that the boundary condition is $E_{\perp,reg}^{(1)}(r_H)=0.$
With this
\be
E_\perp(r)= 1  +  i\wn  \int^r_{C_3}\frac{C}{e^{-\phi}\sqrt{-\gamma}\gamma^{ii}\gamma^{rr}} {\rm d}r  +{\mathcal O}(\wn^2,\qn^2) \, .\label{transol}
\ee
Once the explicit solution has been found, the on shell boundary action can be evaluated  
\be 
S_{B}= - \frac{{\mathcal N}}{2} \int_{r=r_B} {\mathrm d}^{p+1} x\, e^{-\phi} \sqrt{-\gamma} \gamma^{ii} \gamma^{rr} \dA'_{\perp}   \dA_{\perp}\, .
\ee
 and from it, we obtain the correlator  \cite{Son:2002sd} 
\be
 \Pi_\perp(k)=\left.
 {\mathcal N} e^{-\phi} \sqrt{-\gamma} \gamma^{ii} \gamma^{rr} \frac{ E'_{\perp}(r) E_{\perp}(r) }{|E_\perp(r_{B})|^2}
 \right\vert_{r\to r_{B}}\, .
 \label{traspect}
\ee
and we see from (\ref{transol}) that, to this order in $\lambda$, there is no dependence on $\qn$.
Therefore we can write the explicit expression for the conductivity as
\be
\sigma =-\frac{ {1}}{2\pi T}  \lim_{\wn\to 0}  \frac{\im  \Pi_\perp (\wn)}{\wn}\, .
\ee
In fact the imaginary part of (\ref{traspect}) is independent of $r$, and therefore we may evaluate it at the horizon $r \to r_H$ rather than at $r_B$, because there we have explicit control over the singularities. 
Notice that $E_\perp (r) = 1 + {\cal O}(\wn)$, so finally we obtain
\be
\sigma =- {\mathcal N}\, \frac{C}{2\pi T} = 
{\mathcal N}\, e^{-\phi} \sqrt{\gamma \gamma^{00} \gamma^{rr}} \gamma^{ii}\Bigg|_{r\to r_H}\, ,
\label{sigma}
\ee
where use of (\ref{temphawk}) and (\ref{constantC4}) has been made.
For vanishing baryon number $A_t=0$ the matrix $\gamma_{ab}=g_{ab}$ is diagonal and this expression fully agrees with that provided in \cite{Iqbal:2008by}. It generalizes that result to the present context of holographic flavor systems with chemical potential.

\section{The susceptibility \label{secsusc}}

Let us now consider the susceptibility. This is an equilibrium quantity, given by the thermodynamical definition (\ref{defsuscep}). On the one hand we have 
\be
\mu = A_t(r_B) - A_t(r_H)  =  \int_{r_H}^{r_B}  A'_t(r) dr\, ,
\ee
since $A(r_H)=0$ to have a well defined one-form at the horizon (see \cite{Kobayashi:2006sb}). With this
\be
\chi =  \left(\int_{r_H}^{r_B}\frac{d A'_t(r) }{d  n_q}dr 
\right)^{-1} \, .
\label{holochi}
\ee
On the other, from (\ref{betadis}) we can relate the charge density to $A'_t(r)$ 
\be
n_q=  {\cal N}  \frac{e^{-\phi} \sqrt{H} A_t' }{\sqrt{- \left( \gamma_{00}\gamma_{rr}+(2\pi\alpha')^2 A_t'^2 \right)}}\, .
\label{nqat}
\ee 
with $H = \gamma_{ii}^p \gamma_{\theta\theta}^n$. This equation can be inverted for $A_t'$ as a function of $n_q$
\be
A_t' =  \frac{n_q}{{\cal N}}	\sqrt{ \frac{-\gamma_{00}\gamma_{rr}}{\displaystyle e^{-2\phi}H + 
\displaystyle (2\pi\alpha')^2\frac{n_q^2}{{\cal N}^2}}} ~ . 
\label{mudenq}
\ee
Notice that in (\ref{holochi}) we have emphasized the total derivative. This is so because on top of the explicit dependence of $A_t'(r)$ on $n_q$, there is also a hidden one
inside  $\gamma_{rr} = g_{rr}  + \psi'(r)^2g_{\psi\psi}(\psi)$ as well as in the factor  $e^{-2\phi}H = e^{-2\phi}\gamma_{ii}^p \gamma_{\theta\theta}^n$, stemming from the
fact that the brane embedding $\psi(r)$   depends itself parametrically on $n_q$.\footnote{In \cite{Myers:2007we} this fact was overlooked. Remarkably for vanishing baryon number it makes no difference since  the factors multiplying $\d\psi/\d n_q$ and $\d \psi'/\d n_q$ vanish at $n_q=0$.}
In order to simplify  (\ref{mudenq})  it is useful to make use of (\ref{betadis}) and  the following relations
\be
H = \frac{\gamma}{\gamma_{00}\gamma_{rr} + \gamma_{0r}^2}
~~~~;~~~~
n_q^2\frac{(2\pi\alpha')^2}{{\cal N}^2} + e^{-2\phi} H = e^{-2\phi}\gamma \gamma^{00}\gamma^{rr}\, .
\ee
After some algebra one arrives at 
\be
\chi ={\cal N}\left( \int_{r_H}^{r_B} 
\rule{0mm}{6mm}
\frac{1}{e^{-\phi}  \sqrt{-\gamma}\gamma^{00}\gamma^{rr} }
\left[ 
1  + n_q
   \left(\Delta \frac{\d\psi'}{\d n_q}  + \Xi   \frac{\d \psi}{\d n_q} 
\right)
\right]
\right)^{-1}\, .
\label{suscepbaryon}
\ee 
with
\be
\Delta =  \gamma^{rr} \psi'  g_{\psi\psi}  ~~;~~
\Xi = \frac{1}{2}
       \left(\gamma^{rr} \psi'^2 g_{\psi\psi,\psi} - n \gamma^{\theta\theta} \gamma_{\theta\theta,\psi}
   \right)\, .
   \label{DeXi}
\ee
It is remarkable to find the same combinations $\Delta$ and $\Xi$  that  appear naturally in the equations of motion for the fluctuations  (see later in (\ref{olamdos} and  ref. \cite{Mas:2008jz}).

\section{Charge diffusion \label{secdifu} }

In  \cite{Kovtun:2003wp} the diffusion of a generic conserved charge was examined from the
point of view of the membrane paradigm. A closed formula for this quantity was presented, and later
reobtained in the context of the AdS/CFT correspondence in \cite{Starinets:2008fb}. As a warm up exercise, we will rederive this expression in a somewhat simpler way here,  first for $n_q=0$. The expression obtained will also be valid for $n_q\neq 0$ in the case of massless flavors.

\subsection{Zero baryon density}
The charge diffusion constant appears as a pole in the longitudinal  correlator.  In order to find the pole in it,  one must study the equation of motion for the gauge invariant perturbations of the form $E_{||}  = q \dA_{0} + \omega \dA_{1}$.
The relevant equation of motion is 
\be
E_{||}'' + \partial_r \log\left[ \frac{e^{-\phi}\sqrt{-\gamma} \gamma ^{ii} \gamma ^{rr}}{\omega^2  + q^2 \frac{\gamma ^{ii}}{\gamma ^{00}}} \right] E_{||}' - \frac{\omega^2 \gamma ^{00} + q^2 \gamma ^{ii}}{\gamma ^{rr}}E_{||}= 0\, .
  \label{longEOMwithout}
\ee
From the diffusion pole we expect a dispersion relation of the form $\omega = - i \Diff q^2 + ...$. Therefore the natural hydrodynamic scaling is given in terms of a variable $\lambda$ as follows $\omega \to \lambda^2 \omega, \, q\to \lambda q$.
After checking that the indices near the horizon are again $\xi_\pm = 
\pm i \wn/2$, a consistent expansion is
\be
E_{||} (r) = f(r)^{-i\frac{\lambda^2\wn}{2}} \left(E^{(0)}_{||,reg} + \lambda^2 E^{(2)}_{||,reg} + ...\right) = 
E^{(0)}_{||}  + \lambda^2 E^{(2)}_{||}  + ...
\label{scalingsquare}
\ee
Performing a Taylor expansion of the regular part around the horizon $E_{||,reg}(r) = E_{||,reg}(r_H) + 
(r- r_H)E'_{||,reg}(r_H) + ...$ we can obtain $E^{(0)'}_{||}(r_H)$ 
solving the equation of motion (\ref{longEOMwithout}) iteratively 
\be
E_{||}^{(0)}(r_H)=1~,~~~~E^{(0)'}_{||}(r_H) = - \frac{i}{2}\frac{\qn^2}{\wn}  f'(r_H)   .
\label{nearhorexp}
\ee
 Moreover, inserting the ansatz (\ref{scalingsquare}) into the equation of motion (\ref{longEOMwithout}) we may solve at lowest order in closed form
\be
E_{||}^{(0)}(r)=1- i C\frac{\qn^2}{\wn}\int_{r_H}^r\frac{ {\rm d}r }{e^{-\phi}\sqrt{-\gamma}\gamma^{00}\gamma^{rr}} \label{E0long}\, ,
\ee
where the constant in front of the integral was fixed in accordance with (\ref{nearhorexp}) and
(\ref{constantC4}). The relevant Green's function is proportional to 
$\Pi_{||} \sim \lim_{r\to r_{B}} E_{||}'(r)/E_{||}(r_{B})$   where the boundary is at $r_{B}$ \cite{Myers:2007we}.
Hence the dispersion relation comes from demanding that $E_{||}(r_{B}) = 0$.
This provides us with the sought after dispersion relation
$
w  = -i \Diff_0 q^2\, ,
$
with
\be
\Diff_0 = \left.e^{-\phi} \sqrt{\gamma \gamma^{00} \gamma^{rr}} \gamma^{ii}\right\vert_{ r_H} \int_{r_H}^{r_B} \frac{dr}{e^{-\phi} \sqrt{-\gamma}\gamma^{00}\gamma^{rr}}\, .
\label{difuscon}
\ee
Also here,  for vanishing baryon number   $A_t=0 \Rightarrow \gamma_{ab}=g_{ab}$  diagonal and 
(\ref{difuscon})  coincides with the one in \cite{Kovtun:2003wp}.

\subsection{Finite baryon density}

In \cite{Erdmenger:2008yj} a formula close to (\ref{difuscon}) (but not quite) was assumed to
compute the value of $\Diff_0$ also at finite baryon density. The influence of $n_q$ was supposed to show up through the dependence of the coefficients $\gamma_{ab}$ on the profile $\psi(r)$, which itself depends parametrically on the quark density. We shall see in this section that the 
answer that comes from examining the pole of the longitudinal propagator is more subtle.

In the presence of a background value for $A_t$, the longitudinal perturbations examined in the previous section mix with the scalar fluctuations on top of the probe brane profile $\Psi(r)$. The coupled set of equations of motion can be written in the following form
\beqa
E_{||}'' + {\textswab A}_1 E_{||}' + {\textswab B}_1 E_{||} + {\textswab C}_1 \Psi'' + {\textswab D}_1 \Psi' + {\textswab E}_1 \Psi & = & 0\, ,  \label{gauge1} \\
E_{||}'' + {\textswab A}_2 E_{||}' + {\textswab B}_2 E_{||} + {\textswab C}_2 \Psi'' + {\textswab D}_2 \Psi' + {\textswab E}_2 \Psi & = & 0   \, ,\label{gauge2}
\eeqa
where the  explicit form of the coefficients is reproduced from reference   \cite{Mas:2008jz}
in appendix A for completeness.
From there one can easily see that all the coefficients of $\Psi$ are of order $q$. This implies that the natural variable
is $\tilde \Psi = q \Psi$ ($\Psi$ would be natural to go with the gauge potentials $\dA_\mu$).
As before we define $\omega \to \lambda^2 \omega, \, q\to \lambda q$ and it is clear that the coefficients of the terms multiplying $\Psi$ are all of order $\lambda$. Expanding to   order $\lambda^0$, equation (\ref{gauge1}) becomes
  \be 
 E^{(0)''}_{||} + \log'(\sqrt{-\gamma} \gamma^{00}\gamma^{rr}) E^{(0)'}_{||}-
\frac{\gamma^{0r}}{\gamma^{00}\gamma^{rr}}
\left( \Delta \tilde\Psi_{(0)}'  + \Xi \tilde\Psi_{(0)} \right) ' = 0\, ,
\label{olamdos}
\ee
  Using  (\ref{betadis})  this equation can be integrated to give
\be
E_{||}^{(0)} = C_1 + \int_{r_H}^r \frac{\displaystyle C_2 +    n_q\frac{2\pi \alpha'}{\cal N} (\Delta \tilde\Psi_{(0)}' + \Xi \tilde\Psi_{(0)})}{\sqrt{-\gamma}\gamma^{00}\gamma^{rr}}\, ,
\ee
By continuity in the limit  $n_q\to 0$ we may reasonably  expect that the constants $C_1$ and $C_2$ should be the same as in the previous subsection. A more rigorous derivation comes from 
comparing with the Frobenius expansion around, for instance, the horizon, as we did with (\ref{nearhorexp}) and (\ref{E0long}). We have performed this comparison in the specific example of the D3/D7 and found perfect agreement.  Therefore we finally set
\be
E_{||}^{(0)}(r) =1+ \int  \frac{ - i C\displaystyle\frac{\qn^2}{\wn}+   n_q \frac{2\pi \alpha'}{\cal N} (\Delta \tilde\Psi_{(0)}' + \Xi \tilde\Psi_{(0)})}{\sqrt{-\gamma}\gamma^{00}\gamma^{rr}}\, .
\label{eparallel}
\ee
From the Dirichlet boundary condition $E_{||}^{(0)}(r_B)=0$ we obtain the modification to the diffusion constant
\be
\Diff = \frac{D_0}{1+\displaystyle n_q \frac{2\pi \alpha'}{{\cal N}}  \int  \frac{  (\Delta \tilde\Psi_{(0)}' + \Xi \tilde\Psi_{(0)})}{\sqrt{-\gamma}\gamma^{00}\gamma^{rr}}}\, .
\label{diffbaryon}
\ee
The sole purpose of presenting here this rather unwieldy formula for $\Diff$ is to emphasize that the influence of the baryon density on diffusion constant is not just given implicitely by modifying the embedding in   $\Diff_0$ of (\ref{difuscon}). The coupling of the scalar modes that is only present at finite density shows up as an explicit contribution\footnote{A similar coupling can be seen to occur with the vector metric fluctutations in the case of the R-charged black hole. A parallel modification of the membrane-paradigm formula should correctly account for the diffusion at finite R-charge density, (see for example \cite{Son:2006em}).}.
It would be nice to have an expression for $\Diff$ in terms of purely background quantities. 
This would amount to an integration of the equation for the fluctuation $\Psi(r)$ which we have not achieved so far.
However by providing numerical evidence that the Einstein relation holds (see section \ref{secd3d7}), we conjecture that indeed we can express $D$ as $\sigma/\chi$ where both $\sigma$ and $\chi$ are computable in terms of
background quantities $A_\mu(r)$ and $\psi(r)$.

Still an interesting case is that of massless quarks. For such a situation $\Delta$ and $\Xi$ vanish identically, and the diffusion 
constant is given by $D = D_0$ with the dependence on $n_q$ coming through the dependence of
$\gamma_{ab}$  on $\psi(r)$, and in this case the Einstein relation holds with finite baryon density.

\section{Numerics, the D3/D7 case study \label{secd3d7}}
In this section we shall focus on the case of a D3/D7 flavor setup. 
In this particular case we call the radial coordinate $u$, with $u_H=1$ and $u_B=0$. Now $p=n=3,~ \phi=0$ and
\beqa
&& \gamma_{00}=-f(r)\frac{(\pi T L)^2}{u} \hspace{0.5cm} ; \hspace{.5cm} \gamma_{ii}=\frac{(\pi T L)^2}{u}   \hspace{.5cm} ; \hspace{.5cm} \gamma_{0u}=-2\pi\alpha' A_t'(u)  \\
&& \gamma_{uu}=L^2 \frac{1-\psi^2+4u^2f(u)\psi'^2}{4(1-\psi^2) u^2 f(u)}\hspace{.5cm} ; \hspace{.5cm} \gamma_{\theta\theta}=L^2(1-\psi(u)^2) \, ,\\
&& A_t'(u)=-\frac{L^2 T}{4\alpha'}\frac{\Dis\sqrt{1-\psi^2+4u^2f(u)\psi'^2}}{\sqrt{(1-\psi^2)((1-\psi^2)^3+\Dis^2 u^3)}}   \hspace{.5cm} \, ,
\eeqa
where
\be
 f(u)=1-u^2   \hspace{.5cm} ;  \hspace{.5cm} T = \frac{r_H}{\pi L^2} \, .
\ee
Here $\Dis = n_q(2\pi\alpha')/ {\cal N}r_H^3$ is a dimensionless parameter proportional to the density,
whereas the coupling reads
\be
{\mathcal N} = \frac{N_c N_f}{4\pi^2 L^2}\, .\nonumber
\ee
For finite baryon density, only results concerning the conductivity have been reliably established. In \cite{Karch:2007pd} a compact formula was found for the conductivity by implementing holographically 
Ohm's law on a probe D7 brane. A nontrivial check of this expression was provided in 
\cite{Mas:2008jz} by using the spectral functions and the same method explained here in section 
\ref{seccon} 
\be
\sigma=  \frac{N_c N_f T}{4\pi}\sqrt{(1-\psi_0^2)^3 + \Dis^2}\, .
\ee
\begin{figure}[ht]
\begin{center}
\includegraphics[scale=1.7]{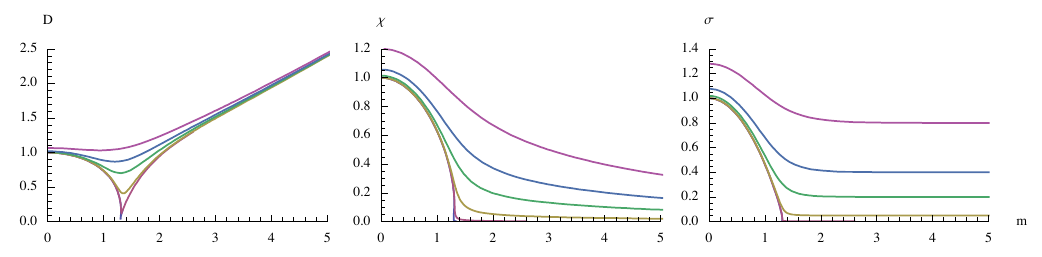}
\caption{\em \label{threeplots}
In this figure we plot several curves for the diffusion constant $\Diff$ (in units of $1/2\pi T$), the susceptibility $\chi$ (in units of $N_c N_f T^2/2$), and the conductivity  $\sigma$ (in units of  $N_cN_f T/4\pi$), against the adimensional mass $m = \bar M/T$. From bottom to top  $\Dis = 0.001,0.005,0.05, 0.2, 0.4$ and $0.8$.
}
\end{center}
\end{figure}
 In figure \ref{threeplots} we show three plots  for the three quantities, $\sigma, \chi$ and $\Diff$, appropriately scaled,  for different values of the baryon density $\Dis = 0.001,0.005,0.05, 0.2, 0.4$ and $0.8$. They are shown as a function of the dimensionless parameter $m = \bar M / T$ with $\bar M = 2M_q/\sqrt{\lambda}$, the mass gap.
  They all behave similarly, approaching  different limiting values for large $m$. In the large $m$ limit, 
  $\sigma$ approaches an $n_q$ dependent constant value whereas $\chi$ dies off as $\sim n_q/m$ and $\Diff$ diverges proportionally to $m$ after  developing a minimum close to the $\Dis = 0$ curve. 
  
  In the massless limit  $m\to 0$ the diffusion constant  $\Diff$ reduces to $ \Diff_0$ in (\ref{difuscon}) which can be expressed in terms of a hypergeometric function  \cite{Kim:2008bv}
\be
D _0= \frac{1}{2 \pi T} \sqrt{1 + \tilde d^2} \phantom \, {}_2F_1
\left(
\frac{1}{3} , \frac{3}{2} ; \frac{4}{3} ; - \tilde d^2 \right)\, . 
\ee

   \begin{figure}[ht]
\begin{center}
\includegraphics[scale=1.4]{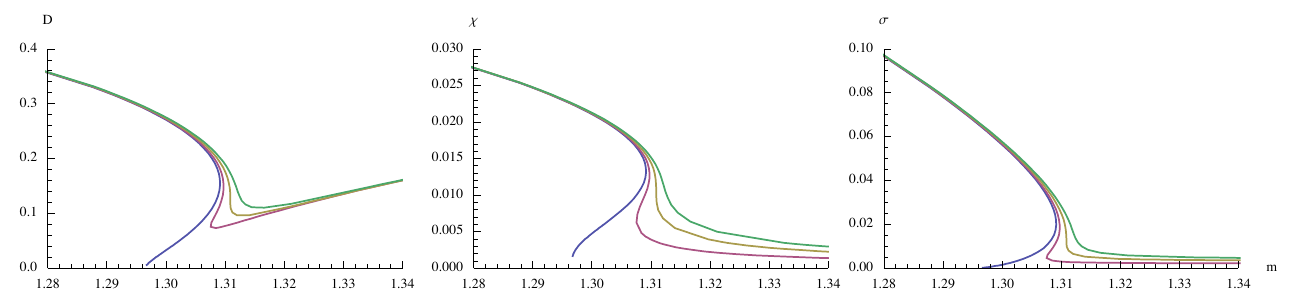}
\caption{\em \label{phases}
 The region close to the end of the $\Dis=0$ (blue) curve is emphasized
 for values of $\Dis = 0.002$, $ 0.00315$ and $0.004$ .  
 }
\end{center}
\end{figure}
In figure \ref{phases} we have zoomed around the region of very small values of $\Dis$.
We observe the  tri-valuedness that all of them exhibit for sufficiently low values of  $\Dis\leq \Dis^* = 0.00315$, which is related to the same property of the embeddings in this region of  the  parameter space
\cite{Kobayashi:2006sb,Erdmenger:2008yj}.

\begin{figure}[ht]
\begin{center}
\includegraphics[scale=1.3]{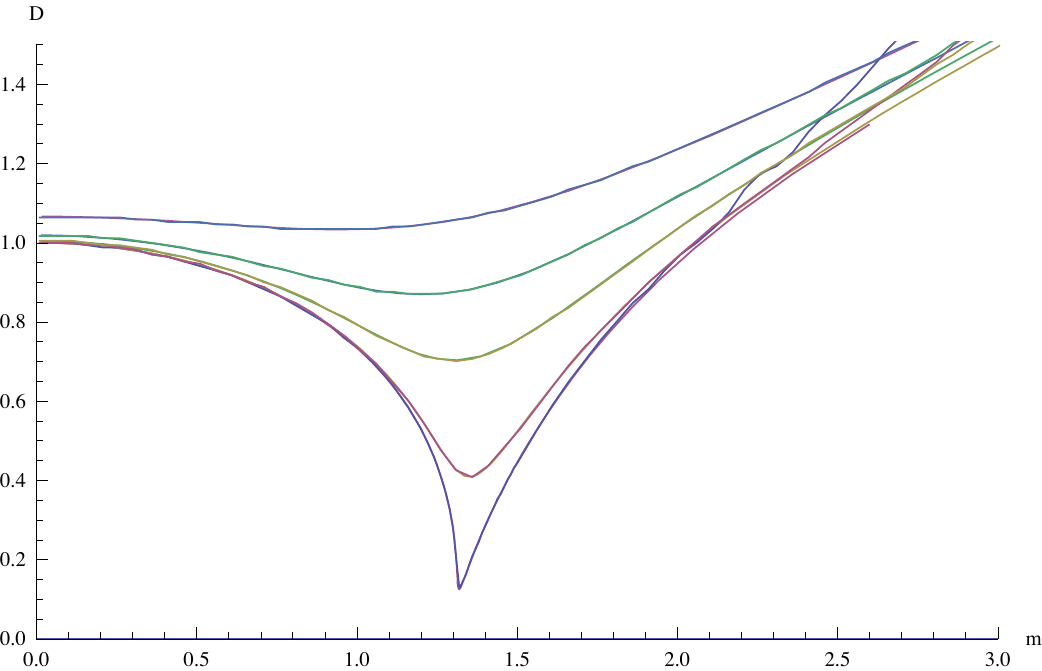}
\caption{\em \label{comparisonplot}
Comparison of the diffusion constant $D$, and the quotient $\sigma/ \chi$. From bottom to top  $\Dis = 0.005,0.05, 0.2, 0.4$ and $0.8$. We see remarkable agreement between the two calculations, up to high values of $\psi_0$ at which point the calculation of the excitations is subject to large numerical instabilities.
 }
\end{center}
\end{figure}
\bigskip
{\bf The Einstein relation}
\medskip

For the case of 
zero baryon number, the analysis performed in \cite{Myers:2007we} confirmed the validity of the
Einstein relation
\be
D  = \frac{\sigma}{\chi} \label{einstrel}\, .
\ee
For nonvanishing baryon number  this equation holds as well in the massless limit $\psi(r)\to 0$. This can be seen from  (\ref{sigma}),(\ref{suscepbaryon}), (\ref{difuscon}) and (\ref{diffbaryon})  after setting $\Xi= \Delta = 0$.
For massive quarks, the validity of (\ref{einstrel}) has to be established numerically. $\Diff$ can be calculated both from a full numerical integration of (\ref{gauge1}) and (\ref{gauge2}) or by integrating 
$\Psi(r)$ and using  (\ref{diffbaryon}). We found complete agreement among them. 
Now, plotting $\Diff$  and $\sigma\chi^{-1}$ leads to the set of curves shown in figure \ref{comparisonplot}. 
We find a very good agreement and the discrepancies arise for high values of $\psi_0 \gtrsim 0.998$ where the numerical computation of $D$ is subject to large instabilities.\footnote{actually for $D$ in figure  \ref{phases}, what we have plotted is the right hand side of (\ref{einstrel}).}
\section{Conclusions}

There is an intimate relationship between the membrane paradigm and the AdS/CFT prescription which is slowly unraveling and getting onto firmer grounds \cite{Starinets:2008fb,Iqbal:2008by,Brustein:2008xx,Fujita:2007fg}.  In this short note we followed the route of the first of these citations and worked fully within the
AdS/CFT context.   In this way we have generalized the closed formula obtained there for the diffusion constant $\Diff$. We have also worked out  the conductivity $\sigma$ and the susceptibility $\chi$. For this last constant  we provide an expression which matches the thermodynamic definition (\ref{defsuscep}). An important technical detail in the calculation of $\chi$ is tracking the implicit dependence of $A_t'$ on $n_q$ through the embedding profile $\psi(r)$ of the flavor brane. Including this contribution  we have shown numerically that, at least for the D3/D7 case,  the three constants obey the Einstein relation (\ref{einstrel}) also at finite baryon number $n_q\neq 0$.  
In the limit of massless quarks we have shown that this relation also holds in general. 

\acknowledgments
We would like to thank  Jorge Casalderrey-Solana, Johanna Erdmenger, Matthias Kaminski, Andreas Karch, Karl Landsteiner, David Mateos, Felix Rust and  Andrei Starinets for correspondence and discussions.
This  work was supported in part by MICINN and  FEDER  under grant
FPA2005-00188,  by the Spanish Consolider-Ingenio 2010 Programme CPAN (CSD2007-00042), by Xunta de Galicia (Conselleria de Educacion and grant PGIDIT06 PXIB206185PR)
and by  the EC Commission under  grant MRTN-CT-2004-005104.  JT has been supported by MICINN of Spain under a grant of the FPU program. JS has been supported by the Juan de la Cierva program.

\begin{appendix}
\section{Coefficients for coupled longitudinal system}
The coefficients that enter the system of differential equations (\ref{gauge1}) and (\ref{gauge2}) 
are reproduced here for completeness (see \cite{Mas:2008jz}).
\beqa
{\textswab A}_1  & = & \log' \left[ \frac{e^{-\phi}\sqrt{-\gamma} \gamma^{ii} \gamma^{rr}}{\omega^2+ q^2 \frac{\gamma^{ii}}{\gamma^{00}}} \right] \, ,\\
{\textswab A}_2  & = &  \log' \left[ e^{-\phi}\sqrt{-\gamma} \gamma^{ii} \gamma^{rr} \right] \frac{\omega^2 \gamma^{00}}{\omega^2 \gamma^{00}+ q^2 \gamma^{ii}} +\frac{\Delta'-\Xi}{\Delta}  \frac{q^2 \gamma^{ii}}{\omega^2 \gamma^{00}+ q^2 \gamma^{ii}}    \, ,\\
{\textswab B}_1  & = & {\textswab B}_2 =  - \frac{\omega^2 \gamma^{00} + q^2 \gamma^{ii}}{\gamma^{rr}} \, ,\\
{\textswab C}_1  & = & -  \frac{q\, \gamma^{0r}}{\gamma^{00}\gamma^{rr}} \Delta   \, ,\\
{\textswab C}_2  & = & - \frac{q(1-\psi' \Delta)}{\psi' \gamma^{0r}}  \, , \\
{\textswab D}_1  & = & -  \frac{q\, \gamma^{0r}}{\gamma^{00}\gamma^{rr}} \left( \Xi + \Delta' \right)  - \frac{q\omega^2}{\omega^2 \gamma^{00} + q^2 \gamma^{ii}}  \frac{\gamma^{0r}}{\gamma^{rr}}  \Delta \,  \log' \left( \frac{\gamma^{ii}}{\gamma^{00}} \right) \, , \\
{\textswab D}_2  & = & - \frac{q(1-\psi' \Delta)}{\psi' \gamma^{0r}} \log'\left[ e^{-\phi}\sqrt{-\gamma} \gamma^{rr} g_{\psi\psi} (1-\psi'\Delta) \right]  \nonumber\\
&& - \frac{\omega^2 \gamma^{00}}{\omega^2 \gamma^{00}+ q^2 \gamma^{ii}}  \frac{q\, \gamma^{0r}}{\gamma^{00} \gamma^{rr}}  \left( \log'\left( e^{-\phi}\sqrt{-\gamma} \gamma^{ii} \gamma^{rr}  \right) - \frac{\Delta'-\Xi}{\Delta} \right) \Delta \, ,  \\
{\textswab E}_1  & = & -  \frac{q\, \gamma^{0r}}{\gamma^{00}\gamma^{rr}} \left( \Xi '  -\Delta \frac{\omega^2 \gamma^{00} + q^2 \gamma^{ii}}{\gamma^{rr}} \right)  - \frac{q\omega^2}{\omega^2 \gamma^{00} + q^2 \gamma^{ii}}  \frac{\gamma^{0r}}{\gamma^{rr}}  \Xi \,  \log' \left( \frac{\gamma^{ii}}{\gamma^{00}} \right) \, , \\
{\textswab E}_2  & = & \frac{q\, \gamma^{0r}}{\gamma^{00}\gamma^{rr}} \left[ \omega^2 \frac{\gamma^{00}}{\gamma^{rr}} \Delta  - \frac{\omega^2 \gamma^{00}}{\omega^2 \gamma^{00}+ q^2 \gamma^{ii}} \, \Xi  \left( \log'\left( e^{-\phi}\sqrt{-\gamma} \gamma^{ii} \gamma^{rr}  \right) - \frac{\Delta'-\Xi}{\Delta} \right) \right]   - \frac{q(1-\psi' \Delta)}{\psi' \gamma^{0r}} H(r) \, , \nonumber\\
\eeqa
with $\Xi$ and $\Delta$ as given in (\ref{DeXi}) and $H$ (not that defined as in the main text) given here
by
\beqa 
H(r)&=&\frac{\partial_r \left(e^{-\phi} \sqrt{-\gamma} \gamma^{rr} \psi' \left( g_{\psi\psi,\psi} + \frac{n}{2} \gamma^{\theta\theta} g_{\theta\theta,\psi}Êg_{\psi\psi} -\med \gamma^{rr}\psi'^2 g_{\psi\psi} g_{\psi\psi,\psi}  \right)  \right)}{e^{-\phi} \sqrt{-\gamma} \gamma^{rr} g_{\psi\psi} (1-\psi'^2 \gamma^{rr} g_{\psi\psi})}
\nonumber\\
&&-\frac{ \left( \omega^2 (\gamma^{00} -  \psi'^2 g_{\psi\psi}(\gamma^{00}\gamma^{rr}+ (\gamma^{0r})^2)) + q^2 \gamma^{ii} (1- \psi'^2 \gamma^{rr}  g_{\psi\psi} )   \right)}{\gamma^{rr}  (1-\psi'^2 \gamma^{rr} g_{\psi\psi})}\nonumber\\
&&-\frac{ \left(  \frac{n(n-2)}{2}  \left( \gamma^{\theta\theta} g_{\theta\theta,\psi} \right)^2 + n \gamma^{rr} \gamma^{\theta\theta} \psi'^2g_{\psi\psi,\psi}Êg_{\theta\theta,\psi}  + n\gamma^{\theta\theta}g_{\theta\theta,\psi\psi} \right)  }{2\gamma^{rr} g_{\psi\psi} (1-\psi'^2 \gamma^{rr} g_{\psi\psi})}\nonumber\\
&& -\frac{\left( \psi'^2 g_{\psi\psi,\psi\psi} -\med \gamma^{rr}\left( g_{\psi\psi,\psi} \right)^2  \psi'^4 \right)  }{2g_{\psi\psi} (1-\psi'^2 \gamma^{rr} g_{\psi\psi})}\, . \label{Hfactor}
\eeqa

\end{appendix}

\end{document}